# One Atomic Beam as a Detector of Classical Harmonic Vibrations with Micro Amplitudes and Low Frequencies

Yong-Yi Huang†

*MOE Key Laboratory for Nonequilibrum Synthesis and Modulation of Condensed Matter,*
*Department of Optic Information Science and Technology,*
*Xi'an Jiaotong University, Xi'an 710049, China*

**Abstract**

We propose a simplest detector of harmonic vibrations with micro amplitudes and low frequencies, i.e. the detector consisting of one atomic beam. Here the atomic beam is induced by a plane harmonic wave and has a classical collective harmonic vibrations, which vibrant directions are perpendicular to the wave vectors of atomic beam. Compared with the detector consisting of atomic Mach-Zehnder interferometer, the new detector has two advantages: (1) it is suitable for the detection of the harmonic vibrations induced either by a longitudinal plane harmonic wave or by a transverse plane harmonic wave; (2) the quantum noise fluctuation of the atomic beam is exactly zero.

†Corresponding author
Email: yyhuang@mail.xjtu.edu.cn

## I. Introduction

Gravitational wave is an extremely weak wave, and it is the only unconfirmed prediction in general relativity. A plane gravitational wave with two polarization states travelling in the z direction will deform the particles around a circle in the xy plane[1], gravitational wave can be confirmed by measuring the oscillatory motions hit by a plane gravitational wave[2] (unfortunately, it is not accepted). Atom interferometer is also designed for the detection of gravitational wave [3,4]. The classical harmonic vibrations are simpler but more fundamental than the oscillatory motions induced by a plane gravitational wave with two polarization states. The effects of atomic beams collective classical transverse harmonic vibrations with micro amplitudes and low frequencies on the mean numbers of atoms arriving at the detectors in atomic Mach-Zehnder interferometer are investigated[5]. The collective classical harmonic vibrations can be detected by measuring the variations of the mean numbers of atoms arriving at the detectors. However two problems are not fully solved in the detector consisting of atomic Mach-Zehnder interferometer: (1) it does not exactly detect the classical harmonic vibrations induced by a transverse plane harmonic wave; (2) there are always quantum noise fluctuations of atomic branches. In this paper we propose a very simple detector consisting of one atomic beam to detect classical harmonic vibrations with micro amplitudes and low frequencies. Although the detector is simple, it is suitable for the detections of the harmonic vibrations induced either by a longitudinal plane harmonic wave or by a transverse plane harmonic wave, the quantum noise fluctuation of the atomic beam is exactly zero. The paper is organized as follows. In section **II** we propose the vibrant factor *F*. In section **III** we present the detection process of the classical harmonic vibrations of micro amplitudes and low frequencies with the new detector consisting of one atomic beam. In section **IV** we discuss the quantum noise fluctuation in the new detector. In section **V** we give a brief summary.

**II. the Vibrant Factor *F***

Our thought experiment schematic is shown in Fig. 1, where there are an atomic oven, a collimator, an atomic beam and a detector. The collimated atomic beam is induced by a plane harmonic wave and is in the same wave front. Just like the detector consisting of Mach-Zehnder interferometer[5], our new detector consisting of one atomic beam can also be used to detect the classical harmonic vibrations induced by a longitudinal plane harmonic wave. As shown in Fig.1, our experimental setup is in the paper plane, we make a longitudinal plane harmonic wave perpendicular to the paper plane hit the atomic beam, the collective harmonic vibrations direction should be perpendicular to the paper plane and perpendicular to the wave vector of the atomic beam. The harmonic vibrant frequencies can be extracted from the number of atoms arriving at the detector. If the harmonic vibrations are induced by a transverse plane harmonic wave, the detector consisting of Mach-Zehnder interferometer can approximately be used to measure the harmonic vibrant frequencies, however, our new detector consisting of one atomic beam can exactly the harmonic vibrant frequencies. What we do is to let a transverse plane harmonic wave in the paper plane be perpendicular to the atomic beam and travel from top to bottom, vice versa. In the following we will discuss the measurement process, one condition should be satisfied, i.e. the collective harmonic vibrant directions are perpendicular to the atomic beam vector, so that the collective vibrations do not couple with translational motions.

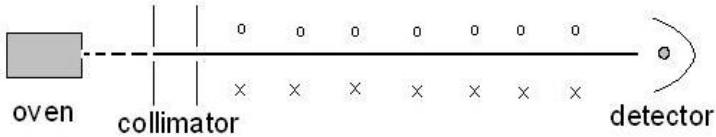

Fig. 1 Schematic of the harmonic vibration detector consisting of one atomic beam, where an atomic beam with certain velocity is produced from an oven with stable temperature. The number of atoms arriving at a detector is measured by a detector after the atomic beam pass through the collimator. The crosses x (perpendicular to the paper plane to the inside) and empty circles o (perpendicular to the paper plane out) of the atomic beam denote the collective classical harmonic vibrations induced either by a longitudinal plane harmonic wave or by a transverse plane harmonic wave. The atomic beam is in the same wave surface and has the same phase. The collective vibrant directions are perpendicular to the wave vectors of the beam.

The atomic beam during the measurement time $\Delta t$ is induced by some plane harmonic wave, which makes the atomic beam collectively vibrate perpendicular to the atomic wave vectors. The collective vibrations of the beam fully decouple with the translational motions of beam. Given the flux of the atomic beam is *j* (the number of atoms per second passing through a plane), the number of atoms in the beam during a measurement time $\Delta t$ is $N = j \cdot \Delta t$. We assume that the measurement time $\Delta t$ is much smaller than the harmonic vibration period and that the efficiency of the detector is unity. Let $|1>_i$ denote the *i*th atom incident on the detector, following ref.[6]

the state vector of the atomic beam with atom number *N* is constructed by a direct product of the individual atomic states, i.e.

$$|\Phi>_N \equiv \prod_{i=1}^{N} |1>_i .  \tag{1}$$

Now we know that the atomic beam has a collective harmonic vibration with the same phase induced by some plane harmonic wave and that the vibrant directions are perpendicular to the wave vectors of the atoms, so the collective vibrations do not couple with translational motions. The Hamiltonian of atomic stream with the number *N* of atoms is given by

$$H = \frac{P_\perp^2}{2M} + \frac{1}{2} M \Omega^2 X^2 + \sum_{i=1}^{N} \frac{p_i^2}{2m},  \tag{2}$$

where $M = Nm$, $m$ is the mass of one atom, and $\Omega$ is the angle frequency of the plane harmonic wave. The state vector of the atomic beam is written as

$$|\Psi>_N = |n> \otimes |\Phi>_N = |n> \otimes \prod_{i=1}^{N} |1>_i  \tag{3}$$

In Equ.(3) $|n>$ is the eigenstate vector of the atomic beam vibration, the corresponding eigen energy is $E_n = (n+1/2)\hbar\Omega$. In this article we assume that $\Omega$ is so low that the behaviors of the harmonic vibration can be regarded as those of a classical harmonic oscillator.

Let $c_i^\dagger$ and $c_i$ be the creation and annihilation operators for the number state $|1>_i$, and we have the number operators $n_i = c_i^\dagger c_i$, where the eigenvalue $n$ is 1. The operator *c* obeys the commutation relationships $c_i c_j^\dagger \pm c_j^\dagger c_i = \delta_{ij}$, where the plus or minus sign indicates Bose or Fermi statistics. However, the statistics are neglected in this article, because we assume that the density of atomic beam is not large and there is only one atom at a time within a single coherence length. The number operator $N$ of the atomic beam is given by

$$N = \sum_{i=1}^{N} n_i .  \tag{4}$$

The expectation value $<N>_N$ of this number operator Equ.(4) is written as

$$<N>_N = {}_N<\Psi|N|\Psi>_N = {}_N<\Phi|\sum_{i=1}^{N} n_i |\Phi>_N \cdot \int_{\Delta t} <n|x><x|n> dx$$
$$= \sum_{i=1}^{N} {}_i<1|n_i|1>_i \cdot \int_{\Delta t} <n|x><x|n> dx = N \int_{\Delta t} <n|x><x|n> dx ,  \tag{5}$$

where $N = j \cdot \Delta t$ has above been declared. Because the measurement time $\Delta t$ is much smaller than the vibration period $T = 2\pi/\Omega$, the mean number of atoms in the detector has a correction $F \equiv \int_{\Delta t} <n|x><x|n> dx$. Here we define the corrected coefficient $F$ of the mean number of atoms arriving at the detectors as the vibrant factor $F$, because it comes from the collective vibrations induced by the plane harmonic wave. The vibrant factor $F$ quantitatively describes the effects of the collective harmonic vibrations on the mean number of atoms in the detector and includes the frequency knowledge of the collective harmonic vibrations. From the vibrant factor $F$, it becomes probable to confirm the existence of the collective vibrations, especially the vibrations with very micro amplitudes, by measuring the variations of the mean number of atoms arriving at the detector.

Under the condition of low frequency, $<n|x><x|n>$ can be regarded as a classical harmonic oscillator probability density. Given $\alpha = \sqrt{M\Omega/\hbar}$ and $\xi = \alpha x$, we obtain the classical vibrant equation of the atomic beam $\xi = \sqrt{2n+1}\sin(\Omega t + \delta)$, where $n$ is the vibrant quantum number, $\delta$ is the initial phase of the atomic beam, and $\sqrt{2n+1}$ denotes the absolute amplitude. The classical harmonic oscillator probability density is $w(\xi) = <n|\xi><\xi|n> = \dfrac{1}{\pi\sqrt{(2n+1)-\xi^2}}$, and $w(\xi)$ increases with the increase of the displacement $\xi$ in $\xi \in [0, \sqrt{2n+1}]$. We do not consider $\xi \in [-\sqrt{2n+1}, 0]$ region because the probability density is symmetric in $[0, \sqrt{2n+1}]$ and $[-\sqrt{2n+1}, 0]$, seen from the probability density $w(\xi) = \dfrac{1}{\pi\sqrt{(2n+1)-\xi^2}}$. During a short measurement time $\Delta t$, we obtain

$$F(\delta) = \int_{\xi_0}^{\xi_0+\zeta} w(\xi)d\xi = \dfrac{1}{\pi}[\arcsin(\dfrac{\zeta}{\sqrt{2n+1}} + \sin\delta) - \delta], \qquad (6)$$

where $\zeta$ denotes the absolute displacement during the measurement time $\Delta t$, and $\delta$ is the initial phase i.e. $\xi_0 = \sqrt{2n+1}\sin\delta$ with $\xi_0$ being the initial displacement at the initial time $t_0 = 0$. Equ. (6) is a fundamental formula in this article, which was firstly derived in [7]. It is the condition the measurement time $\Delta t << T$ that makes the integral upper limit and lower limit in Equ.(6) be $[\xi_0, \xi_0 + \zeta]$ rather than $(-\infty, +\infty)$ and the relative displacement be small i.e.

$0 < \dfrac{\zeta}{\sqrt{2n+1}} << 1$.

The factor $F$ versus the initial phase $\delta$ is shown in Fig. 2 given the relative displacement

$\zeta/\sqrt{2n+1} = 0.05$. Here $\zeta/\sqrt{2n+1} = 0.05$ is not a particular choice but is just an example, the maximum of the relative displacement is unity. Actually we study the vibrant factor $F$ during one period T and require the measurement time $\Delta t$ be less than one period T. Seen from Fig.2 and Equ.(6) we observe: (1) The vibrant factor $F$ depends on the relative vibrant displacement $\frac{\zeta}{\sqrt{2n+1}}$ and initial phase $\delta$, rather than the absolute vibrant amplitude $\sqrt{2n+1}$. (2) The vibrant factor $F$ increases with the increase of the initial phase $\delta$. The reason is that the different initial phases $\delta$ correspond to the different displacements $\xi$, and the probability density $w(\xi)$ increases with the increase of displacement $\xi$.

During the measurement time $\Delta t << T$ the small relative displacement $\frac{\zeta}{\sqrt{2n+1}}$ is much smaller than unity, and the integral upper limit and lower limit in Equ.(6) is $[\xi_0, \xi_0 + \zeta]$ rather than $(-\infty, +\infty)$, so the vibrant factor $F$ is less than unity. What's more, the vibrant factor $F$ reduces the mean number of atoms in the detector by two orders of magnitude. For instance, given $\frac{\zeta}{\sqrt{2n+1}} = 0.05$, we have $\sin\delta = 0.95$, i.e. $\delta \simeq 1.25$ maximally from the fact $\frac{\zeta}{\sqrt{2n+1}} + \sin\delta = 1$. Substituting $\delta \simeq 1.25$ into $F(\delta)$, we obtain that $F(\delta)$ maximum is about 0.1. If $\frac{\zeta}{\sqrt{2n+1}} = 1$ is satisfied in Equ.(6), we have $\sin\delta = 0$. The maximal value of $F(\delta)$ is 0.5, which corresponds to one-half period. In a whole period we have $F(\delta) = 1$, however, it is a trivial result because we can not extract useful information.

**III. Detection of the Classical Harmonic Vibrations of Micro Amplitudes and Low Frequencies with the Detector Consisting of One Atomic Beam**

Given $\frac{\zeta}{\sqrt{2n+1}} = 0.05$ in the detection process of number of atoms, the mean number of atoms arriving at the detector is given by

$$_N<\Psi|N|\Psi>_N = N \cdot F(\delta), \tag{7}$$

where $F(\delta) = \frac{1}{\pi}[\arcsin(0.05 + \sin\delta) - \delta]$ and $N = j \cdot \Delta t$. Seen from Fig. 2, the larger the initial phase $\delta$ becomes, the longer measurement time $\Delta t$ it takes to keep the same relative displacement $\zeta/\sqrt{2n+1}$ to be constant. Maybe it is difficult and not practical to use this method to measure the mean number of atoms arriving at the detector during a short measurement time $\Delta t$, however, theoretically once the left curve $F$ versus the phase is obtained, the classical collective vibrations of atomic beam are verified. How does one obtain the vibrant factor $F$? Without the collective classical harmonic vibrations, the mean number of atoms arriving at the detector is written as

$$<\Phi|\sum_{i=1}^{N}n_i|\Phi>_N = \sum_{i=1}^{N}{}_i<1|n_i|1>_i = N. \tag{8}$$

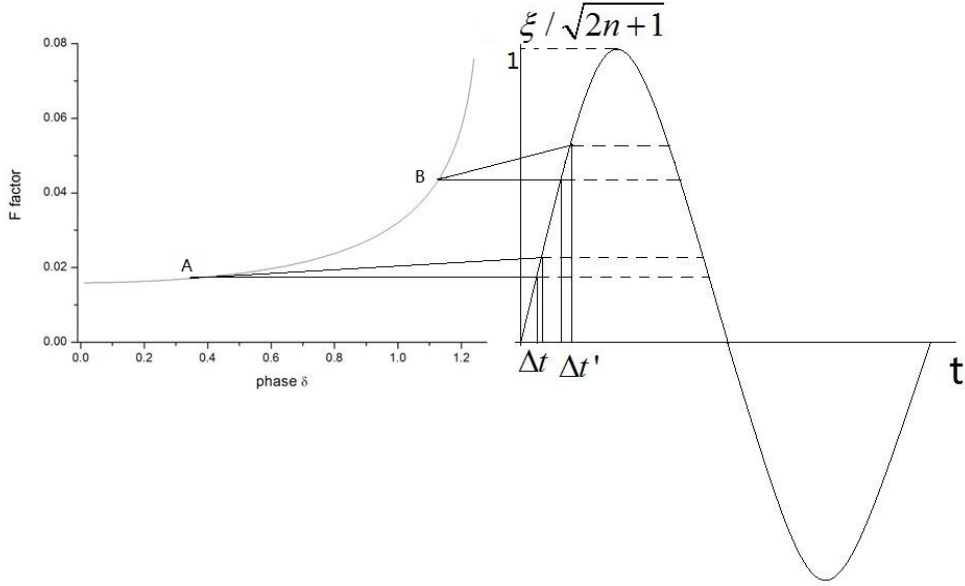

Fig. 2 Vibrant factor $F$ versus the initial phase $\delta$ with the relative displacement $\zeta/\sqrt{2n+1}=0.05$ lies in the left panel, and the right panel corresponding to the vibrant factor $F$ is the time-varying harmonic vibrant displacement with the same frequency $\Omega$, reproduced from ref.[5]. A dot in the left panel corresponds to a measurement time $\Delta t$ with an initial phase $\delta$, and it takes different measurement time $\Delta t$ at different phases $\delta$ to keep the relative displacement $\zeta/\sqrt{2n+1}$ to be constant. For instance, dot A and dot B in the left panel have different initial phases and correspond to different displacements in the right panel. If the relative displacement $\zeta/\sqrt{2n+1}$ is kept to be constant (the measurement time in dot A is less than the time in dot B i.e. $\Delta t < \Delta t'$), the vibrant factor $F(A)$ is less than $F(B)$ because of the probability in dot A less than the probability in dot B seen from Equ.(6).

It is obvious that we can get the curve of the vibrant factor $F$ after comparing the modified results Equ.(7) with Equ.(8).

The other practical measurement of the mean number of atoms is equal time interval measurement, i.e. we keep the measurement time $\Delta t$ to be constant. As shown in Fig.2, a relative vibrant displacement $\frac{\zeta}{\sqrt{2n+1}}$ contains two to-and-fro processes in one-half period, so we need to divide the vibrant factor $F(\delta)$ in Equ.(6) by 2 in a practical measurement process. Let the measurement time $\Delta t$ be constant, we surprisedly obtain $\frac{F(\delta)}{2} = \frac{\Delta t}{T}$ independent of the initial phase $\delta$. The mean number of atoms arriving at the detector is given by

$$_N<\Psi|N|\Psi>_N = j \cdot \Delta t \cdot \frac{\Delta t}{T}. \tag{9}$$

As for the multi-period measurement, we have to replace $\Delta t$ by $m\Delta t$, $T$ by $mT$, where $m$ is the number of periods. Seen from Equ.(9), the vibrant factor corrections are invariant and independent of the number of measurement periods. It is obvious that we can evaluate the frequency $T$ of the atomic beam after comparing the modified results Equ.(9) with Equ.(8).

**IV. The Quantum Noise Fluctuation of the Atomic Beam**

Without considering the harmonic vibrations of the two atomic branches, when the beams intensity is so low that there is only one atom at a time within a single coherence length, the quantum noise fluctuations in the detector consisting of atomic Mach-Zehnder interferometer are given by[6]

$$<\Delta N_{A,B}>_0 = \frac{\sqrt{N}}{2}\sin\varphi_{\alpha\beta}, \tag{10}$$

where we have $N = j \cdot \Delta t$ and $\varphi_{\alpha\beta} \equiv k(l_\alpha - l_\beta)$ with $k$ being the atomic wave vector, $l_\alpha, l_\beta$ being the path lengths through the $\alpha$ and $\beta$ branches. Given $\varphi_{\alpha\beta} \neq 0$ there always exist the quantum noise fluctuations in the two detectors of Mach-Zehnder interferometer.

If the quantum noise fluctuations of the beams are larger than the number of atoms arriving at detectors during a small measurement time $\Delta t$, the number of atoms arriving at detectors will not be stable, it is difficult to measure the accurate frequency of the beams. The thought setup needs a stable beam with a very small even zero quantum noise fluctuations, which is very difficult to be realized in the detector consisting of atomic Mach-Zehnder interferometer. We will illustrate that there is exactly zero quantum noise fluctuation in the new detector consisting of one atomic beam. Without the classical harmonic vibrations the quantum noise fluctuation in the beam is given by

$$\begin{aligned}<\Delta N^2> &= <\Phi|N^2|\Phi>_N - (<\Phi|N|\Phi>_N)^2 \\ &= <\Phi|\sum_{i=1}^N n_i \sum_{i=j}^N n_j|\Phi>_N - (<\Phi|\sum_{i=1}^N n_i|\Phi>_N)^2 \\ &= \sum_{i=1}^N <1|n_i|1>_i \sum_{\substack{j=1 \\ j\neq i}}^N <1|n_j|1>_j + \sum_{i=1}^N <1|n_i^2|1>_i - N^2 \\ &= N(N-1) + N - N^2 = 0\end{aligned} \tag{11}$$

Because of the existence of the zero quantum noise fluctuation of the atomic beam, we do not consider the effects of the quantum noise fluctuations on the number of atoms arriving at the detector. So there are not limits about the measurement time $\Delta t$. Due to the quantum noise fluctuations in the detector consisting of atomic Mach-Zehnder interferometer, the measurement number of atoms arriving at detectors during very small time $\Delta t$ may be submerged in the quantum noise fluctuations of the branches.

**V. Summary**

In conclusion a simplest detector of harmonic vibrations with micro amplitudes and low frequencies, i.e. the detector consisting of one atomic beam is proposed. The new detector has two advantages: (1) it is suitable for the detections of the classical harmonic vibrations induced either by both a longitudinal plane harmonic wave or by a transverse plane harmonic wave; (2) the quantum noise fluctuation of the atomic beam is exactly zero. The new detector is simpler but more practical than the detector consisting of atomic Mach-Zehnder interferometer. The present results maybe provide a new and probable detection principle for gravitational wave.

What about the realistic detector for the classical harmonic vibrations? The detection efficiency for atoms should be very high, best close to unity. If the thought setup is further improved to detect the vibrations induced by gravitational wave, we speculate that the setup is not very large, and that it should be in vacuum and be laid in a satellite to receive the vibrations induced by gravitational wave.